\documentclass[graybox]{svmult}

\usepackage{type1cm}        
\usepackage{makeidx}         
\usepackage{graphicx}        
\usepackage{multicol}        
\usepackage[bottom]{footmisc}

\usepackage{newtxtext}       %
\usepackage{newtxmath}       

\makeindex             

\begin{document}

\title*{Game-Theoretic Foundations for  Cyber Resilience Against Deceptive Information Attacks in Intelligent Transportation Systems}
\titlerunning{Game-Theoretic Foundations for Cyber
Resilience in ITS}
\author{Ya-Ting Yang and Quanyan Zhu}
\institute{Ya-Ting Yang \at New York University, \email{yy4348@nyu.edu}
\and Quanyan Zhu \at New York University, \email{qz494@nyu.edu}}
%
%
\maketitle

\abstract{The growing complexity and interconnectivity of Intelligent Transportation Systems (ITS) make them increasingly vulnerable to advanced cyber threats, particularly deceptive information attacks. These sophisticated threats exploit vulnerabilities to manipulate data integrity and decision-making processes through techniques such as data poisoning, spoofing, and phishing. They target multiple ITS domains, including intra-vehicle systems, inter-vehicle communications, transportation infrastructure, and human interactions, creating cascading effects across the ecosystem. This chapter introduces a game-theoretic framework, enhanced by control and learning theories, to systematically analyze and mitigate these risks. By modeling the strategic interactions among attackers, users, and system operators, the framework facilitates comprehensive risk assessment and the design of adaptive, scalable resilience mechanisms. A prime example of this approach is the Proactive Risk Assessment and Mitigation of Misinformed Demand Attacks (PRADA) system, which integrates trust mechanisms, dynamic learning processes, and multi-layered defense strategies to counteract deceptive attacks on navigational recommendation systems. In addition, the chapter explores the broader applicability of these methodologies to address various ITS threats, including spoofing, Advanced Persistent Threats (APTs), and denial-of-service attacks. It highlights cross-domain resilience strategies, offering actionable insights to bolster the security, reliability, and adaptability of ITS. By providing a robust game-theoretic foundation, this work advances the development of comprehensive solutions to the evolving challenges in ITS cybersecurity.}

\section{Introduction}
The increasing complexity and interdependence of Intelligent Transportation Systems (ITS) make them particularly vulnerable to cyber threats. Modern ITS integrates a network of technologies, such as automated vehicles \cite{lei2022cognitive}, smart traffic management \cite{li2024digital}, and communication systems \cite{tutorial2018game}, to ensure seamless operation. However, this interconnectedness means that a cyber attack on one component can ripple across the entire system \cite{mecheva2020cybersecurity}. For example, an attack on a traffic control center can disrupt communications, traffic flow, and logistics across a region. Addressing these inter-dependencies through robust cyber resilience measures is essential to ensure the continuous and safe operation of ITS \cite{pan2023resilience}.

ITS faces a dynamic and growing array of cyber threats. Attackers exploit vulnerabilities in transportation systems, using ransomware to lock systems, denial-of-service (DoS) attacks to overwhelm networks, and unauthorized access to compromise sensitive data or take control of systems \cite{usama2024cyber}. Such attacks are not limited to opportunistic hackers; state-sponsored actors and organized crime groups also target critical transportation infrastructure. To combat these threats, ITS must employ advanced threat detection tools, real-time monitoring systems, and proactive defense mechanisms to identify and neutralize potential risks before they escalate.

The cybersecurity of transportation systems has direct implications for public safety. Cyber attacks on automated vehicles, traffic signals, or transportation control systems can lead to accidents, injuries, or even fatalities. For instance, a compromised automated vehicle system could malfunction, endangering passengers and other road users. Similarly, a breach in emergency response networks could delay critical assistance during accidents or disasters. Ensuring cyber resilience \cite{zhu2024foundations} is not merely a technical challenge but a public safety priority, critical to maintaining trust in transportation systems and protecting lives.

The economic impact of cyber attacks on ITS can be significant. Transportation systems are vital to global commerce, enabling the movement of goods and people. A cyber incident could disrupt supply chains, delay deliveries, and lead to financial losses for businesses and consumers \cite{cartwright2023economics}. For example, ransomware attacks on freight logistics systems can halt deliveries and impact industries reliant on timely supply chains. By building resilience into ITS, agencies can prevent such economic disruptions and ensure that transportation continues to support economic stability and growth.

The adoption of advanced ITS technologies, such as automated and connected vehicles, depends heavily on public trust. Users need to feel confident that these technologies are safe and reliable \cite{zhang2024public}. A high-profile cyber attack that compromises safety or disrupts operations can erode public trust, slowing the adoption of these innovations. Cyber resilience is therefore essential not only for preventing attacks but also for demonstrating a commitment to security, fostering confidence among users, and encouraging the widespread adoption of advanced transportation technologies.

With the growing emphasis on cybersecurity, regulatory bodies are mandating stringent standards for transportation systems. Frameworks like the NIST Cybersecurity Framework \cite{NIST.CSWP.29} and strategies like zero-trust architectures \cite{ge2023zero,stafford2020zero} require transportation agencies to adopt comprehensive security practices. Compliance with these standards not only protects systems from cyber threats but also aligns transportation infrastructure with national and international security objectives. By adhering to these frameworks, ITS operators can ensure a consistent and proactive approach to security.

Addressing the cybersecurity challenges of ITS requires a detailed understanding of attack vectors and their implications. In the first part of the chapter, we delve into deceptive information attacks, highlighting their mechanisms and potential consequences for ITS. These attacks manipulate data integrity and the decision-making processes in ITS, utilizing techniques like data poisoning, spoofing, and phishing, which exploit the system and human vulnerabilities \cite{buinevich2019forecasting}. For example, data poisoning and injection attacks compromise data servers that collect operational inputs, resulting in flawed decisions by machine learning tools. These attacks skew models, leading to unsafe or inefficient system operations. Similarly, jamming and spoofing target physical signals, affecting vehicular control, potentially causing road accidents by transmitting false GPS coordinates or disrupting V2V (vehicle-to-vehicle) and V2I (vehicle-to-infrastructure) communications. Phishing attacks, often part of sophisticated Advanced Persistent Threats (APTs) \cite{huang2020dynamic}, exploit human vulnerabilities. Such attacks may lead to the installation of malware that traverses both IT and OT networks, compromising critical assets like trains, traffic signals, or control systems, showcasing the intersection between human error and technical vulnerabilities.

To understand this, we create multi-domains, including intra-vehicle, inter-vehicle, transportation, and human domains. Each domain presents unique challenges and is susceptible to specific types of attacks that exploit its vulnerabilities. The intra-vehicle domain focuses on the internal systems of a vehicle, including control units, sensors, and communication modules. Attacks in this domain, such as malware, ransomware, and adversarial machine learning, target critical functionalities like braking, steering, and navigation. These attacks often exploit software vulnerabilities or user actions, such as connecting to unsecured networks or falling victim to phishing scams, potentially compromising the safety and reliability of the vehicle.
The inter-vehicle domain involves communication between vehicles and between vehicles and infrastructure. This domain is particularly vulnerable to data injection, spoofing, and man-in-the-middle attacks, which exploit protocols like V2V (vehicle-to-vehicle) and V2I (vehicle-to-infrastructure). These attacks can lead to significant disruptions, including traffic congestion, false routing information, and unsafe driving conditions.

The transportation domain encompasses the broader infrastructure that supports ITS, such as traffic management centers, traffic signals, and dynamic pricing systems. Attacks in this domain, including denial-of-service (DoS), data manipulation, and advanced persistent threats (APTs), can disrupt operations on a large scale, creating traffic chaos, financial losses, and risks to public safety by compromising critical systems.

The human domain involves the interaction between humans—such as drivers, operators, and users—and ITS. Attacks in this domain often exploit human vulnerabilities through social engineering, phishing, and cognitive manipulation \cite{huang2023cognitive}. For example, misleading drivers with falsified traffic updates or tricking operators into deploying malicious software can propagate malware or lead to unsafe decision-making.

The domains are interdependent, and attackers often exploit vulnerabilities across multiple domains to achieve their goals. For instance, a phishing attack targeting an operator in the human domain could introduce malware into the transportation domain, which may propagate to the inter-vehicle domain and ultimately compromise the intra-vehicle domain. A domain-specific view is essential for understanding and mitigating targeted attacks within a single domain, while a cross-domain view is crucial for identifying cascading effects and interdependencies. Both perspectives are vital to achieving the overarching goal of cybersecurity and resilience in ITS.

In the second part of the chapter, we evaluate and quantify the risks associated with deceptive information attacks in ITS, providing a foundation for designing resilience mechanisms to mitigate their impact. A central approach to achieving this involves leveraging game theory \cite{zhu2019game}, complemented by tools from control and learning theories \cite{zhu2024foundations}. Together, these methodologies offer a structured framework for both assessing risks and crafting adaptive, robust defenses.

Game theory provides a systematic way to model the strategic interactions between attackers and defenders, enabling a deeper understanding of adversarial behaviors and optimal response strategies. It allows for the analysis of various scenarios, such as resource allocation for defense, the propagation of misinformation, and the cascading effects of attacks across interconnected domains. For example, Stackelberg games \cite{zhao2024stackelberg,li2024meta,yang2023game} can model leader-follower dynamics where attackers initiate malicious actions, and defenders adaptively respond to minimize disruptions. Control theory complements this by offering tools to stabilize systems under attack, ensuring operational continuity and preventing cascading failures. By integrating dynamic models, control mechanisms can anticipate potential system instabilities caused by attacks and implement corrective measures in real time. Techniques such as adaptive control and fault-tolerant control play a critical role in maintaining system resilience.

Learning theories, particularly reinforcement learning \cite{nguyen2021deep}, further enhance resilience by enabling ITS systems to learn from historical and real-time data. These tools can identify evolving attack patterns and adjust defense strategies accordingly. For instance, learning algorithms can refine game-theoretic models by incorporating new insights, enabling ITS to adapt dynamically to emerging threats. The integration of machine learning with game theory \cite{fang2021introduction} also supports predictive risk modeling, allowing systems to anticipate adversarial behaviors and proactively deploy mitigation measures. By combining these disciplines, ITS can move beyond static defenses to implement adaptive and cross-layer resilience mechanisms. These mechanisms not only respond to ongoing attacks but also evolve to address future threats, ensuring the robustness and safety of intelligent transportation systems in the face of complex and interdependent cyber risks.

The organization of the paper is as follows. In Section 2, we introduce the key domains of ITS and the specific challenges and vulnerabilities they face, including intra-vehicle, inter-vehicle, transportation, and human domains. This section lays the groundwork by describing the interdependencies among these domains and how attackers exploit them to achieve their goals. In Section 3, we present game theory as a foundational framework for analyzing and designing resilience strategies. This section discusses how game-theoretic models, combined with control and learning theories, can be used to assess risks, predict adversarial behaviors, and develop adaptive defense mechanisms to enhance ITS resilience. Key concepts such as Stackelberg games, dynamic games \cite{chen2021dynamic,chen2021dynamic_}, and signaling \cite{pawlick2018modeling,li2022commitment} games are highlighted as tools to model and counter deceptive information attacks.

In Section 4, we use a case study to demonstrate the practical application of these theoretical frameworks. The case study focuses on navigating the risks of misinformed demand attacks on ITS navigational systems, showcasing how game-theoretic models and resilience mechanisms can mitigate cascading failures, maintain system stability, and address both local and system-wide impacts of such attacks. This section also provides insights into the effectiveness of proposed strategies in real-world scenarios. The chapter concludes with Section 5, where we discuss the future directions on the deceptive information threats and the resilience mechanisms to protect ITS. 

\section{Deceptive Information Attacks}
This chapter focuses on deceptive information attacks within ITS, which are designed to create misinformation \cite{yang2023strategic}, manipulate data \cite{pan2022poisoned}, and compromise system integrity through techniques like data poisoning and injection. These attacks can leverage malware (e.g., advanced persistent threats or APTs), man-in-the-middle tactics, or jamming and spoofing. They can occur across multiple layers of the ITS, which can be broadly categorized into intra-vehicle, inter-vehicle, and transportation system domains, as well as human interactions.

\subsection{Intra-Vehicle Domain}

Vehicles, as complex control systems, are increasingly vulnerable to a variety of cyber threats, including malware, spoofing, and man-in-the-middle attacks. These threats can compromise the safety and reliability of critical vehicle functions, particularly in autonomous and connected vehicles. For instance, adversarial machine learning attacks can manipulate or poison the training data used by AI perception systems, leading to incorrect object detection or decision-making, which could undermine the functionality of features like lane-keeping, collision avoidance, or traffic sign recognition \cite{pan2023stochastic,chi2024adversarial}. Additionally, ransomware attacks represent a growing concern, especially as vehicles frequently connect to external networks for software updates or infotainment services. These attacks can encrypt vehicle systems or data, rendering them inoperable, and are often facilitated by user negligence, such as falling victim to phishing attempts or connecting to unsecured networks.

An example of malware targeting vehicles is the ``Jeep Cherokee hack" \cite{king_2016} demonstrated by cybersecurity researchers in 2015. The attack exploited vulnerabilities in the vehicle's Uconnect infotainment system, allowing remote access to critical controls such as the engine, brakes, and steering. The researchers were able to wirelessly send malicious commands through the vehicle's cellular connection, showing how an attacker can disable the brakes or take control of steering while the vehicle is in motion. This event highlighted the real-world dangers of malware attacks on vehicles and emphasized the urgent need for improved cybersecurity in the automotive industry.

To enhance resilience in the intra-vehicle domain, robust mechanisms must be implemented. Cloud-enabled backup control systems \cite{zhu2020new}, such as those explored in \cite{zhu2020secure,chen2017security,pawlick2018istrict}, provide a failsafe by allowing vehicles to revert to a secure and functional state even in the event of a cyber attack. Furthermore, deploying layered protections for vehicle control systems is critical. This includes encrypting communication channels, implementing strong authentication protocols, and employing intrusion detection systems to identify and mitigate threats in real-time.

In addition to technical safeguards, organizational measures such as cyber insurance \cite{liu2023cyber} and accountability frameworks \cite{ge2022accountability} are essential for mitigating the financial and operational impacts of attacks. Cyber insurance can help offset the costs associated with recovery and liability, while clear accountability measures ensure that stakeholders prioritize cybersecurity best practices, from vehicle manufacturers to end-users. Together, these strategies form a comprehensive approach to protecting vehicles from cyber threats, safeguarding their functionality, and maintaining user trust in modern automotive technologies.

\begin{figure}
\includegraphics[width=3.5in]{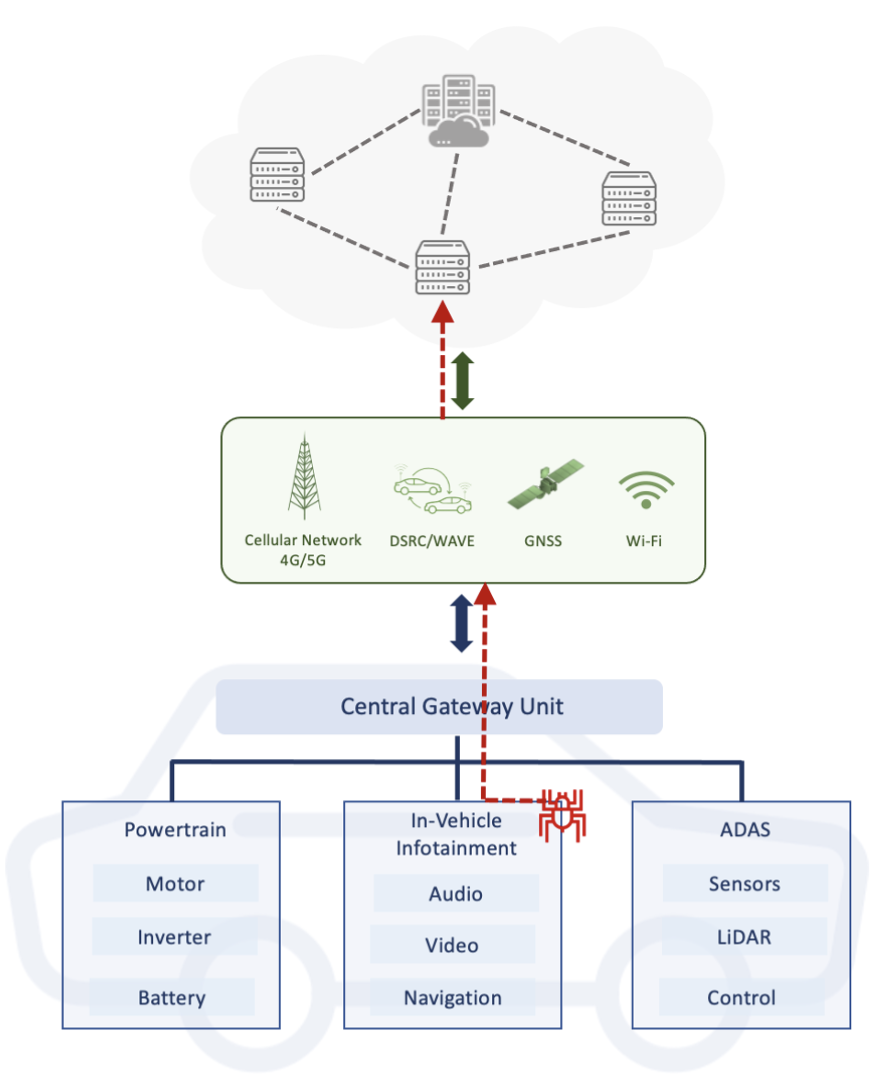}
\centering
\caption{Intra-vehicle domain attacks: An attacker can exploit vulnerabilities in the in-vehicle infotainment system, such as audio, video, or navigation modules, to compromise the central gateway unit. Once the gateway is breached, the attack can propagate through vehicle-to-infrastructure (V2I) communications using technologies like cellular networks (4G/5G), DSRC/WAVE, GNSS, or Wi-Fi, potentially impacting cloud servers and disrupting critical services across the connected vehicle ecosystem}
\label{fig:intra_v}
\end{figure}

\subsection{Inter-Vehicle Domain Attacks}

The inter-vehicle domain encompasses the communication infrastructure that supports cloud and edge computing, intersection management, and interference mitigation. This domain involves a vast network of nodes that are highly dynamic, with nodes joining and leaving frequently. Such dynamics increase risks, including opportunities for ransomware propagation and disruptions due to infrastructure interdependencies.

The adoption of 5G technologies introduces additional security concerns. There is a need for robust measures such as zero-trust architectures for edge computing services at intersections These measures must manage access control, establish trust, and meet quality of service (QoS) requirements. Without such safeguards, the inter-vehicle domain remains highly susceptible to data manipulation and other informational attacks. 

To address these vulnerabilities, zero-trust architectures \cite{ge2022trust} are essential for securing the inter-vehicle domain. This approach operates on the principle of ``never trust, always verify'', treating every entity within the network as potentially hostile. Key measures include fine-grained access controls, such as multi-factor authentication and continuous monitoring, to ensure that only authorized entities can access network resources. Trust can be further enhanced using blockchain-based identity verification \cite{liu2020blockchain} to prevent impersonation and spoofing attacks. Additionally, ensuring QoS is vital to prioritize critical applications, such as emergency vehicle coordination, and mitigate threats like denial-of-service (DoS) attacks.

Without these safeguards, the inter-vehicle domain remains vulnerable to a range of informational attacks, including data injection, eavesdropping, and manipulation. For example, attackers could inject false data into vehicle-to-infrastructure (V2I) communication \cite{gupta2020secure}, causing traffic congestion or unsafe driving conditions. Similarly, man-in-the-middle (MITM) attacks \cite{al2020review} could intercept and alter messages between vehicles, leading to accidents or service disruptions.

One example is Simon Weckert, who tricked Google Maps into creating a fake traffic jam in Berlin \cite{Weckert2020}. The attack involved Weckert pulling a wagon loaded with 99 smartphones, all running Google Maps, through a street in Berlin. The system interpreted the cluster of devices as heavy vehicular traffic, resulting in Google Maps rerouting drivers to avoid the area. While this was a benign demonstration, it highlighted the vulnerabilities in ITS to data manipulation attacks. Furthermore, in \cite{eryonucu2022sybil}, researchers demonstrated a Sybil attack on Google Maps, which created a significant impact on traffic management and user experience within intelligent transportation systems (ITS). By generating numerous fake identities, the attacker manipulated Google Maps to display fabricated congestion, leading to misinformed traffic routing. 

\begin{figure}
\includegraphics[width=3.5in]{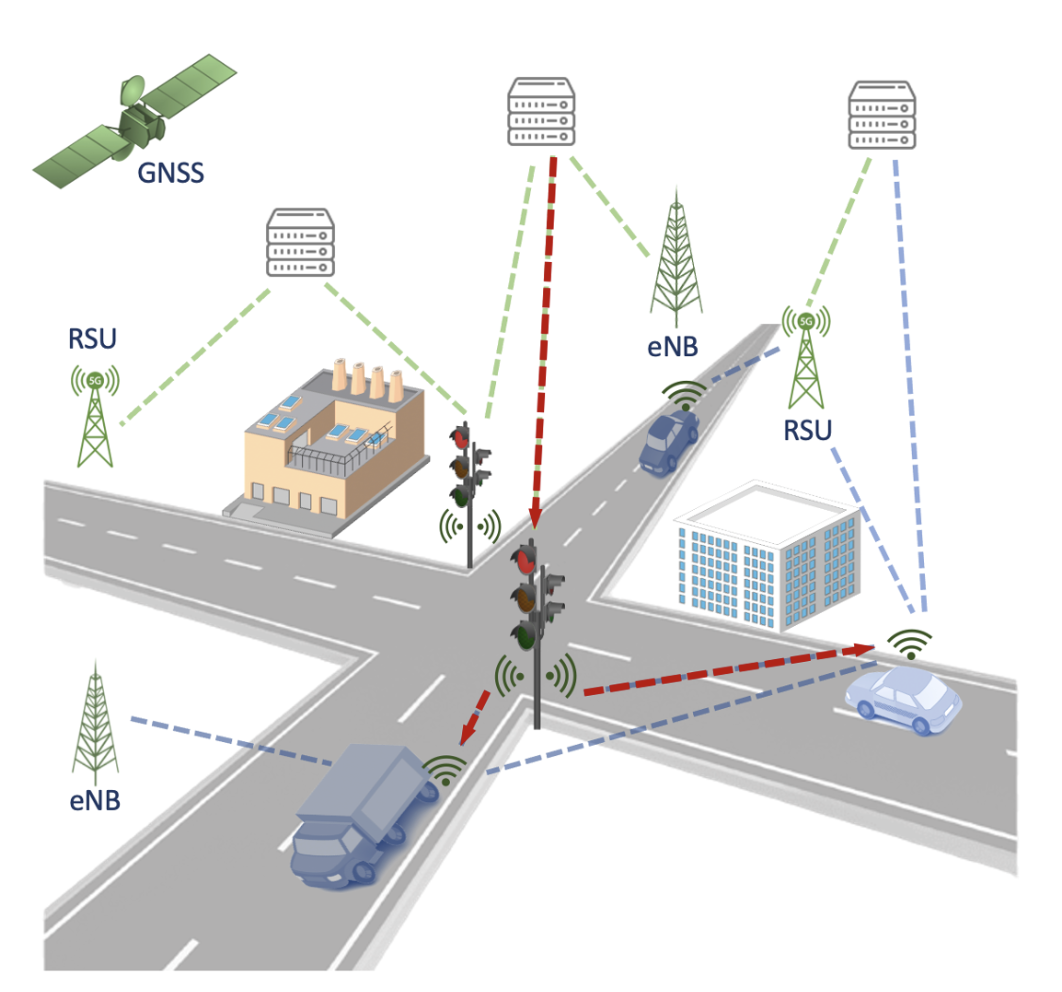}
\centering
\caption{Attacks in the Inter-Vehicle Domain: Smart transportation systems rely on the integration of GNSS, roadside units (RSUs), and enhanced Node BS (eNBs) to facilitate seamless vehicle-to-infrastructure communication and optimize traffic flow at urban intersections. However, these systems are vulnerable to deceptive information attacks. For instance, an attacker can compromise an edge server, poisoning the traffic data it collects. This misinformation can then be propagated to vehicles, leading to incorrect routing decisions and the deliberate creation of traffic congestion.}
\label{fig:inter_v}
\end{figure}

\subsection{Transportation System Domain}
\begin{figure}
\includegraphics[width=0.99\linewidth]{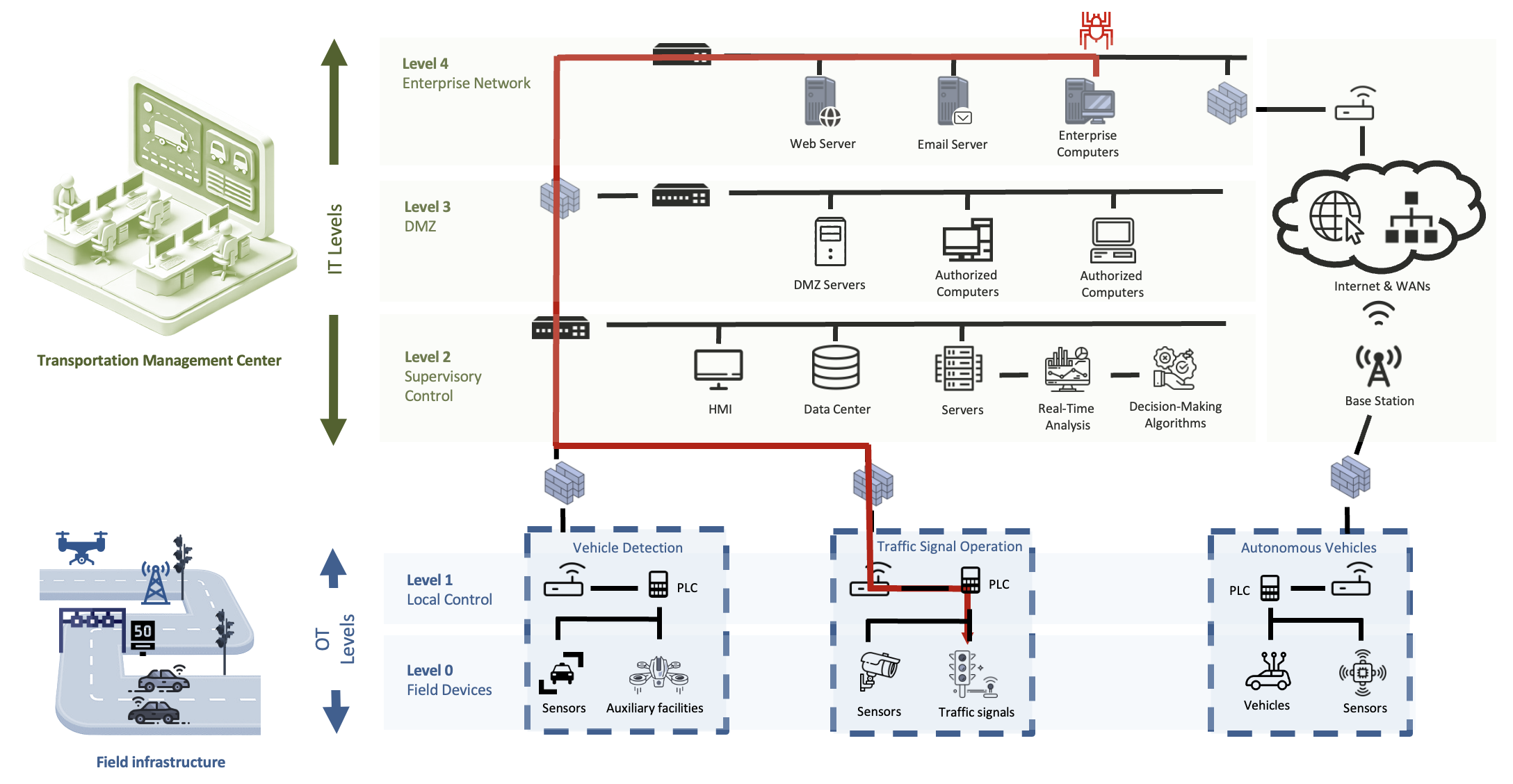}
\centering
\caption{Attacks in the Transportation Domain: Malware can infiltrate the enterprise network of a transportation control center, often originating from a compromised workstation through phishing, malicious downloads, or other attack vectors. Once inside, the malware can corrupt or poison critical databases that store and process real-time traffic data. This can lead to falsified decision-making, such as generating inaccurate traffic flow recommendations or manipulating vehicle routing algorithms, thereby causing congestion or unsafe road conditions. Additionally, the attack can escalate to the OT level, where the malware targets interconnected physical assets such as traffic light signaling systems, toll collection points, or electronic message boards. By compromising these systems, attackers can manipulate traffic lights to cause accidents, disrupt toll collection to create revenue loss, or display misleading information to drivers, amplifying the overall impact on road safety and traffic management. The cascading effects of such attacks can result in significant economic and social disruption.}
\label{fig:trans_domain}
\end{figure}

The transportation system domain is a pivotal layer of ITS, including traffic management systems and centers that house critical infrastructure such as traffic control mechanisms, dynamic pricing systems, and other essential operational components. Unlike the inter-vehicle domain, which emphasizes communication between vehicles and between vehicles and road infrastructure, the transportation system domain focuses on the overarching operation and control of the transportation network from a supervisory perspective. Figure \ref{fig:trans_domain} provides a detailed overview of the layered architecture that connects IT and OT systems in a Transportation Management Center (TMC). The framework is designed to integrate high-level enterprise management with field-level operational control, ensuring the secure and efficient management of modern transportation systems.

At the highest level, the IT system encompasses enterprise and supervisory functions. Level 4 (Enterprise Network)includes critical systems like web servers, email servers, and enterprise computers, which support administrative tasks, decision-making, and overall management of the transportation infrastructure. Below this, Level 3 (DMZ) serves as a security layer, isolating sensitive enterprise systems from operational networks. It houses DMZ servers and authorized computers to mediate secure interactions between IT and OT domains. Moving further down, Level 2 (Supervisory Control) facilitates operational oversight through tools like Human-Machine Interfaces (HMIs), data centers, real-time analysis modules, and decision-making algorithms, enabling effective control of transportation operations.

The OT system focuses on field-level control and interaction with physical infrastructure. Level 1 (Local Control) comprises Programmable Logic Controllers (PLCs) and similar devices that handle localized operations, such as vehicle detection, traffic signal management, and autonomous vehicle coordination. These controllers translate supervisory commands into actionable outputs at the field level. At the base, Level 0 (Field Devices) includes sensors and auxiliary facilities that interact directly with the physical environment, monitoring conditions like traffic flow, vehicle positions, and signal status to feed real-time data back into the system.

There are two-way communications between IT and OT systems for the hierarchical flow of information and control. Secure gateways ensure that high-level decisions made within the TMC translate into actionable directives for field-level systems. A red path within the diagram underscores potential vulnerabilities, drawing attention to cybersecurity risks that could compromise system functionality and safety.

Attacks can occur at multiple levels within a transportation management system, posing significant risks to both IT and OT components. At Level 0, field devices, such as sensors and controllers, are particularly vulnerable to compromise. Attackers can manipulate these devices to disrupt their functionality, leading to inaccuracies in vehicle detection or traffic signal operation. These disruptions can cascade, undermining the broader transportation network.

In addition to direct field-level attacks, cyber threats from the IT domain pose even greater risks. Cyberattacks, including malware, ransomware, and APTs, can infiltrate the enterprise network of a transportation control center, as illustrated in Figure \ref{fig:trans_domain}. Often, these attacks originate from compromised workstations via phishing campaigns, malicious downloads, or other sophisticated attack vectors. Once inside the system, the malware can corrupt or poison critical databases that store and process real-time traffic data. This can result in falsified decision-making, such as inaccurate traffic flow recommendations or manipulated vehicle routing algorithms, which can cause severe congestion, unsafe road conditions, or even system-wide failures.

The threat does not stop at the enterprise network; it can escalate to the OT level, where malware targets interconnected physical assets. These include traffic light signaling systems, toll collection points, and electronic message boards. By exploiting these systems, attackers can disrupt real-time operations, manipulate physical infrastructure, and exacerbate the impact of the initial compromise.

The consequences of such attacks can extend far beyond localized disruptions, with the potential to cripple entire cities. For example, a compromised traffic control system can lead to widespread gridlock by altering traffic light timing or creating unsafe intersections \cite{feng2022cybersecurity}. Similarly, malware targeting train signaling systems can cause delays, operational shutdowns, or even collisions \cite{unger2023securing}. Furthermore, breaches in payment and pricing systems can result in stolen financial data or disrupted revenue streams for metropolitan transit authorities, compounding the economic and operational impacts \cite{huq2017cyberattacks}.

These vulnerabilities indicate the critical importance of securing operational technologies within ITS. Ensuring the resilience and security of OT systems is essential to maintaining the reliability and safety of modern transportation networks, protecting public safety, and supporting uninterrupted mobility and commerce. OT within ITS plays a foundational role in ensuring the seamless operation of critical transportation infrastructure. Unlike traditional IT, which primarily focuses on data processing and storage, OT systems are designed to control and monitor physical processes, such as traffic signal operations, train signaling, and road sensor functionality. Given their critical role in transportation safety and efficiency, OT systems must not only be secure but also resilient, i.e., capable of maintaining functionality and recovering quickly in the face of threats or failures \cite{rieger2009resilient,rieger2019industrial,rieger2012agent}.

OT resilience goes beyond conventional cybersecurity measures, as it must account for the operational continuity and physical safety of systems \cite{ishii2022security,rieger2019industrial,zhu2011robust}. For example, while IT systems prioritize protecting data confidentiality, OT systems focus on ensuring the availability and reliability of operational processes. Any disruption to OT, whether from a cyber attack or a physical fault, can have immediate and far-reaching impacts, such as traffic congestion, delayed emergency responses, and increased safety risks. Therefore, resilience in OT systems means equipping them with mechanisms to prevent, detect, and respond to attacks while maintaining critical operations \cite{zhu2024disentangling}. The following gives two examples of resilience to deceptive information attacks.

(a) Resilience of Traffic Conditions: OT systems must ensure that traffic flows remain stable and efficient even when under attack. Consider a scenario where a misinformation attack provides falsified traffic data to navigation systems \cite{ning2023robust}, misleading drivers to congest specific routes unnecessarily. Resilience mechanisms such as dynamic traffic management algorithms, rooted in principles like Wardrop equilibrium, can mitigate these impacts by redistributing traffic loads in real-time, ensuring optimal usage of available road capacity \cite{correa2011wardrop}. Similarly, in the case of data manipulation attacks targeting sensors, systems can leverage redundant data sources or anomaly detection algorithms to validate and correct manipulated inputs.

(b) Resilience of Traffic Control Systems: Traffic control systems, including pricing mechanisms and signaling systems, are particularly vulnerable to data injection attacks \cite{almalki2021deep}. For instance, attackers could inject false pricing data into toll-collection systems \cite{jolfaei2023survey}, either disrupting revenue collection or causing confusion among drivers. Resilient systems can counteract such attacks by using advanced techniques to verify the integrity of data and applying real-time monitoring to detect anomalies \cite{zhu2020cross,xu2015secure}. Similarly, signaling systems, such as those used for train operations, must maintain strict operational integrity. A resilient system would isolate compromised signals, revert to predefined safety configurations, and notify operators to prevent collisions or service disruptions, maintaining safety and efficiency under duress.

\subsection{Human Aspects}
Humans play a pivotal role in the ITS ecosystem, not only as users but also as decision-makers, and their responses to information—accurate or otherwise—significantly influence traffic conditions. Social media and other communication platforms can amplify the impact of misinformation \cite{yang2023designing}, giving rise to information-induced DoS attacks. For example, falsified traffic recommendations or panic-inducing misinformation can lead to erratic driver behaviors, unnecessary rerouting, or even traffic jams, further exacerbating the vulnerabilities of the ITS infrastructure.

Both users and operators are susceptible to deceptive informational attacks. Users can be misled into downloading malicious software, which may compromise intra-vehicle systems. Through inter-vehicle communications, this malware can propagate across the entire transportation network, creating a systemic risk. Similarly, operators at transportation management centers are vulnerable to social engineering attacks that trick them into deploying malware. Once inside, such malware could disrupt key operations, alter traffic signals, or corrupt routing algorithms, leading to widespread congestion and unsafe road conditions.

These attacks often exploit cognitive vulnerabilities inherent in human decision-making \cite{huang2024psyborg}. One such vulnerability is base rate neglect, where individuals fail to consider the base rate or statistical probability of an event when making decisions. For example, in transportation networks, a driver might overreact to misinformation about an accident ahead, ignoring the statistical improbability of a major incident affecting multiple routes simultaneously. This could result in unnecessary detours or panic-induced congestion, compounding traffic issues rather than alleviating them.

Another common cognitive vulnerability is confirmation bias \cite{katakwar2023attackers}, where individuals selectively seek or prioritize information that aligns with their preexisting beliefs, ignoring contradictory evidence. For instance, a user predisposed to distrust automated navigation systems might disregard accurate routing suggestions during an emergency, relying instead on unverified social media updates. This can lead to suboptimal route choices, further straining the transportation network.

By exploiting these cognitive biases, attackers can manipulate both users and operators, leading to cascading effects that disrupt the ITS and create widespread chaos. One recent example of such exploitation occurred during a cyber-induced traffic manipulation attack, where attackers disseminated falsified traffic alerts through social media platforms and GPS navigation apps \cite{waniek2021traffic}. Users, influenced by confirmation bias, trusted these updates because they aligned with their assumptions about typical traffic congestion during peak hours. As a result, many drivers followed incorrect detour routes, which led to significant congestion in previously uncongested areas, amplifying delays, and frustrating commuters.

Another instance involved base rate neglect during a coordinated misinformation campaign targeting transportation management centers \cite{cox2020stuck}. Attackers tricked operators into believing a high-probability cyber event, such as a widespread signal outage, was underway. They achieved this by using phishing emails that mimicked alerts from legitimate monitoring systems. The operators, failing to consider the statistical unlikelihood of such an outage, deployed emergency protocols that unnecessarily rerouted traffic and temporarily disabled interconnected systems. This not only disrupted traffic flows but also created vulnerabilities for additional attacks.

Such examples highlight how cognitive biases are not merely abstract vulnerabilities but can be directly weaponized by attackers to cause real-world consequences. These manipulations can cripple ITS infrastructure by overloading resources, undermining trust in automated systems, and creating feedback loops of inefficiency. Mitigating these risks requires integrating cognitive and behavioral insights into ITS design and cybersecurity protocols, ensuring that systems are resilient to both technical attacks and human-induced errors.

Several works have focused on modeling cognitive biases and mitigating their adversarial exploitation, particularly in the context of human-centric systems such as ITS and at a broader level Human-Cyber-Physical Systems (HCPS). These efforts often integrate insights from cognitive science, psychology, and system science to create defenses against attacks that exploit human vulnerabilities. For instance, reactive and proactive attention attacks have been extensively studied, with models developed to counteract the impact of Informational Denial-of-Service (IDoS) attacks, where feint alerts induce cognitive overload \cite{huang2023advert,huang2023radams}. Techniques such as Q-learning via consolidated data and visual-aid designs have been proposed to enhance attention and improve operator decision-making. Such mechanisms aim to filter critical signals from noise, reducing the likelihood of manipulation through cognitive overload.

Other approaches leverage game theory and decision-theory models to predict and counteract adversarial strategies that exploit biases like base rate neglect \cite{cox2020stuck} and confirmation bias \cite{katakwar2023attackers}. For example, by incorporating principles of Cumulative Prospect Theory \cite{tversky1992advances}, defenses can be tailored to address how individuals over- or under-weigh probabilities during decision-making, which attackers often exploit. Furthermore, human-centered AI technologies have been developed to mitigate memory vulnerabilities and cognitive manipulation. Examples include human-machine interfaces (HMI) that adapt to user cognitive states, providing explainable recommendations and real-time feedback based on biosensor inputs like heart rate or motion signals. These systems aim to minimize errors caused by biases such as anchoring or framing by offering contextual explanations and alerts during critical decision-making.

\section{Cross-Layer Resilience}
Resilience in ITS cannot be effectively ensured by focusing on individual components or isolated layers. The interconnected nature of ITS—spanning vehicles, transportation infrastructure, and human interactions—demands a \textbf{holistic systems approach} that recognizes and addresses the interplay between these layers. Each layer depends on the others to function optimally, and vulnerabilities in one layer can propagate across the system, leading to cascading failures that impact the entire transportation ecosystem.

For instance, consider a scenario where malware infiltrates a single vehicle through a compromised onboard system, such as an infotainment or telematics application. While the initial infection might appear localized, the interconnected nature of ITS means that this malware can exploit vehicle-to-vehicle (V2V) and vehicle-to-infrastructure (V2I) communication channels to spread. This could disrupt real-time traffic management systems, disable critical infrastructure like traffic signals, or compromise the coordination of autonomous vehicles. The ripple effects may include system-wide traffic congestion, delayed emergency response times, increased safety risks, and significant economic consequences for individuals and businesses relying on transportation services.

The interplay between ITS layers amplifies risks, as failures in one area can trigger disruptions in others. For example, human behavior, such as negligence or lack of cybersecurity awareness, is often the initial trigger for attacks. A driver might inadvertently download a malicious app or click on a phishing link, introducing malware into a vehicle’s system. This seemingly isolated event could escalate if the malware exploits vulnerabilities in the V2I ecosystem, infecting traffic control systems and causing widespread disruptions. Similarly, compromised infrastructure components like road sensors or edge computing nodes could relay false data to vehicles, causing unsafe driving conditions or inefficient traffic flows.

These interdependencies underscore the importance of addressing resilience at a system-wide level. Ensuring that each layer is secure and resilient on its own is insufficient; the connections and interactions between layers must also be safeguarded.

\subsection{Holistic Framework for Resilience}
To effectively protect ITS from cyber threats and operational disruptions, a comprehensive resilience framework must address vulnerabilities across all layers of the system. This framework ensures seamless coordination between components and layers, enhancing the system's ability to prevent, detect, and recover from potential failures or attacks.

\subsubsection{Layer-Specific Defenses}
Resilience begins with tailored security measures for each layer of ITS, addressing the unique challenges and requirements of vehicles, infrastructure, and human interactions. Modern vehicles, increasingly dependent on sophisticated software and communication systems, are prime targets for cyber attacks. Robust defenses include IDS to monitor vehicle networks for unauthorized access, encrypted communication protocols to secure data exchanges, and regular software updates to address vulnerabilities. Fail-safe mechanisms are also critical, enabling vehicles to revert to safe operational modes in the event of a system compromise.

Infrastructure components such as traffic management systems, edge computing nodes, and communication networks form the backbone of ITS. Securing these elements involves network segmentation to isolate critical infrastructure from non-essential systems, multi-factor authentication, and role-based access control, and anomaly detection tools to identify and respond to potential threats.

Human interaction is another crucial layer, as human error remains a significant vulnerability in ITS. Cybersecurity education and awareness campaigns are essential to train users and operators on best practices, such as recognizing phishing attempts and securing connected devices. Additionally, user-centric design \cite{huang2022advert} can minimize errors caused by complex interfaces or cognitive overload.

\subsubsection{Cross-Layer Coordination}
ITS resilience relies on the ability of different layers to work together seamlessly. Effective coordination ensures that a threat detected in one layer is communicated across the system for a unified response. For example, if a vehicle detects abnormal behavior, such as malware activity, it should notify nearby vehicles and infrastructure systems. This allows the broader system to isolate the threat, reroute traffic, or take other protective actions. Shared intelligence platforms where vehicles, infrastructure, and operators exchange real-time threat data can further enhance coordinated responses. Ensuring interoperability across devices and systems, regardless of vendor or protocol, is also critical for maintaining resilience.

\subsubsection{Proactive Threat Mitigation and Cascading Failure Containment}
Anticipating and addressing vulnerabilities before they are exploited is a cornerstone of resilience. Predictive analytics can leverage historical data and real-time monitoring to identify patterns indicative of emerging threats, such as unusual congestion caused by misinformation attacks. Machine learning algorithms can detect subtle anomalies, such as a slow-moving APT, enabling timely interventions. Regular risk assessments \cite{zhang2019mathtt} ensure that the ITS adapts to evolving technologies and threat landscapes, identifying and addressing weak points proactively.

One of the greatest risks in ITS is the potential for a failure in one component to propagate across the system. Containment strategies are vital to isolate and manage such failures effectively. Redundancy in critical infrastructure, such as duplicate traffic control servers or alternative communication paths, ensures operational continuity during failures. Fail-safe systems are designed to automatically revert to a secure and functional state when compromised. For example, a compromised traffic sensor can be isolated while redundant inputs maintain accurate traffic flow predictions. Similarly, compromised autonomous vehicles can switch to manual operation or predefined safe-driving modes. Layer isolation ensures that issues within one layer, such as a compromised vehicle network, do not cascade into others like traffic management systems.

Consider a scenario where malware infects a vehicle's system through a malicious app. The vehicle's IDS detects the anomaly and isolates the infected components. Simultaneously, it sends alerts to nearby vehicles and infrastructure systems, prompting a coordinated response. Traffic management systems reroute unaffected vehicles to reduce congestion caused by the compromised vehicle. Anomaly detection tools monitor the network for further signs of malware propagation, while predictive algorithms assess the likelihood of similar attacks elsewhere. These coordinated efforts contain the impact, prevent escalation, and ensure continued system functionality.

\subsection{Theoretical Foundations and Design Frameworks}
Game theory \cite{fang2021introduction,manshaei2013game} offers a robust mathematical framework for understanding and designing interactions among agents in ITS, particularly under conditions of uncertainty, adversarial actions, and competing interests. This framework is essential for creating distributed resilience mechanisms that enable ITS components—such as vehicles, infrastructure systems, and users—to make optimal decisions that balance individual goals with the overall system's objectives.

Game theory captures the strategic dynamics between different entities in ITS, such as defenders (system operators) and attackers (cyber adversaries), as well as between cooperative and non-cooperative agents within the system. These interactions often involve adversarial scenarios where attackers aim to exploit vulnerabilities, and defenders must anticipate and mitigate potential disruptions. By modeling these interactions, game theory provides insights into optimal strategies for both attack and defense, helping system designers create more resilient frameworks \cite{abdallah2024game,zhu2015game}.

For example, autonomous vehicles might collaborate to share information about road conditions to enhance safety and efficiency. Cooperative game theory helps determine how resources, such as bandwidth or computational power, are allocated among vehicles to maximize collective resilience. In scenarios where agents have competing objectives, such as prioritizing emergency vehicle routing versus minimizing congestion for general traffic, non-cooperative game theory helps balance these conflicting interests. Given the inherent uncertainty in ITS environments, stochastic games can model dynamic scenarios where agents adjust their strategies based on evolving conditions, such as unpredictable traffic patterns or emerging cyber threats.

\subsubsection{Dynamic Games for Cyber Resilience}
The complexity of ITS requires dynamic game models that adapt to changes in real-time. Dynamic games, as discussed in \cite{zhu2024foundations}, involve sequential interactions where the actions of one stage influence the strategies and outcomes of subsequent stages. For instance, in \textbf{proactive resilience}, defenders might reconfigure traffic management systems to reduce vulnerabilities before an attack occurs. During a \textbf{responsive phase}, the system might employ adaptive mechanisms like rerouting traffic to mitigate the impact of a detected attack. \textbf{Retrospectively}, game theory supports analyzing attacker strategies to refine defensive measures and prevent future incidents.

Dynamic games are particularly valuable for modeling cyber resilience, as they account for the evolving tactics of attackers and the adaptive responses of defenders. These models provide a structured way to plan for multi-stage interactions to ensure that ITS remains resilient under dynamic and adversarial conditions \cite{zhu2012dynamic,chen2019dynamic,nugraha2020dynamic}.

\subsubsection{Leveraging Asymmetric Information and Learning}
In many ITS scenarios, agents operate under asymmetric information—where one party (e.g., attackers) has more knowledge about vulnerabilities or system states than others (e.g., defenders). Game-theoretic frameworks such as Bayesian games and signaling games are particularly effective for addressing these challenges.
For instance, in a \textbf{signaling game} \cite{pawlick2019game}, attackers may send deceptive signals to mislead ITS components, such as providing false traffic data to divert vehicles or disrupt traffic flow. Game theory helps design mechanisms that enable defenders to distinguish genuine information from false signals, enhancing system resilience. Similarly, asymmetric information games are instrumental in designing \textbf{zero-trust architectures} \cite{ge2023scenario,li2024decision}, where the system continuously verifies user behaviors to dynamically assess trustworthiness.

\subsubsection{Integration with Learning Frameworks}
Game theory becomes even more powerful when combined with learning mechanisms. As attackers evolve their tactics, ITS systems can employ reinforcement learning to adapt their strategies in real time, continuously refining game-theoretic models with updated data \cite{huang2022reinforcement}. This integration offers several advantages. One is Predictive Modeling, where ITS can anticipate adversarial behaviors based on historical patterns and dynamic interactions, enabling proactive defense measures. The other one is Continuous Adaptation. Systems can dynamically adjust defenses to counter emerging threats, such as new types of malware targeting connected vehicles. By combining game-theoretic models with machine learning \cite{kamhoua2021game}, ITS resilience frameworks can operate with greater flexibility and precision, addressing both known and unknown threats effectively.

Game-theoretic models play a crucial role in addressing the complex challenges within Intelligent Transportation Systems (ITS). These models are particularly valuable in cyber-physical security, traffic management, and resource allocation, enabling strategic decision-making that enhances resilience and efficiency across the system.

In \textbf{cyber-physical security}, game-theoretic models are used to analyze and mitigate coordinated attacks on ITS infrastructure. These attacks often exploit vulnerabilities in both cyber and physical layers, such as malware targeting traffic management servers or tampering with field devices like traffic lights. Game theory provides a framework for anticipating attacker strategies and designing optimal defensive measures. For instance, it can guide the deployment of decoy systems to mislead attackers or recommend the isolation of compromised nodes to prevent malware spread \cite{pawlick2021game}. Additionally, game-theoretic models help mitigate cascading failures by predicting how disruptions in one part of the system could impact interconnected components \cite{zhu2012dynamic}, thereby informing strategies to contain the damage and maintain overall functionality.

In \textbf{traffic management} \cite{li2024multi}, game-theoretic frameworks address conflicting objectives, such as minimizing delays for general traffic while prioritizing emergency vehicle passage. Cooperative game theory facilitates collaboration between traffic lights and autonomous vehicles to optimize signal timings, reducing congestion and improving overall flow. In contrast, non-cooperative game theory is useful when different agents, like emergency responders and civilian drivers, have competing priorities. These models help determine equitable strategies that balance the system's needs without compromising safety or efficiency. By modeling these interactions, game theory ensures that traffic management systems can operate effectively even under complex, dynamic conditions.

Effective \textbf{resource allocation} is another area where game theory excels \cite{yang2022edge}. ITS relies on limited resources, such as communication bandwidth, computational power, and energy, to function optimally. Game-theoretic models enable agents to make strategic decisions about resource use, ensuring critical operations are prioritized. For example, autonomous vehicles and traffic management nodes can negotiate bandwidth allocation based on urgency, ensuring collision avoidance data is transmitted with minimal latency. Similarly, computational resources at edge nodes can be allocated to high-priority tasks, such as processing real-time traffic data, while less critical operations are deferred. These strategies optimize resource utilization and maintain system reliability.

\subsection{Benefits of Game-Theoretic and Learning-Based Design Principles for Cyber Resilience}
The integration of game-theoretic models and learning-based approaches into ITS resilience frameworks provides several key benefits, particularly when these frameworks adhere to foundational design principles such as scalability, interoperability, real-time responsiveness, and adaptability.

Scalability is essential for ensuring that resilience mechanisms can handle the increasing complexity and scale of ITS. Game-theoretic models are inherently suited for distributed solutions, where individual components make localized decisions that collectively scale across the network. For example, a decentralized traffic management system based on cooperative games can efficiently expand from managing a single intersection to coordinating traffic across an entire metropolitan area. This scalability allows ITS to grow without compromising resilience.

Interoperability is critical in ITS, which encompasses diverse devices, protocols, and data sources. Resilience frameworks must integrate these heterogeneous components seamlessly \cite{zhao2024stackelberg,zhao2023stackelberg,zhao2023stackelbergIFAC,zhao2024learning}. Game-theoretic strategies provide a flexible framework for collaboration among agents with varying capabilities, such as autonomous vehicles from different manufacturers. These strategies enable interoperability by establishing common rules and protocols for interaction, ensuring that the overall system functions cohesively despite technical diversity.

The real-time responsiveness of resilience mechanisms is vital for addressing dynamic threats and disruptions in ITS. Learning-based frameworks, such as reinforcement learning \cite{li2024symbiotic,hammar2024automated,li2022sampling}, complement game-theoretic models by enabling agents to adapt strategies as conditions evolve. For example, reinforcement learning can update game-theoretic models in response to sudden congestion caused by an accident or a cyber attack, ensuring that defenses remain effective. This combination of real-time adaptability and strategic foresight enhances the system's ability to manage unexpected challenges.

Finally, robustness and adaptability are core principles of cyber resilience. Robustness ensures that ITS can withstand known threats, while adaptability allows it to respond effectively to novel challenges \cite{zhu2024disentangling}. A dual-layer approach that combines rule-based strategies with adaptive learning is particularly effective. Predefined game-theoretic models handle predictable scenarios, such as routine resource allocation, while machine learning algorithms refine these models to counter emerging threats. For example, if attackers develop new malware targeting V2V communication, reinforcement learning can update defense strategies to mitigate the risk \cite{li2023self,li2022sampling,ge2023scenario}.

\section{Case Study}

\subsection{Misinformation attacks on recommendation systems}

One practical case study for applying resilience frameworks in ITS by utilizing game-theoretic methods is the Proactive Risk Assessment and Mitigation of Misinformed Demand Attacks (PRADA) framework. PRADA, \cite{yang2024prada}, addresses vulnerabilities such as those shown in Fig. \ref{fig:intra_v}, \ref{fig:inter_v}, \ref{fig:trans_domain} for navigational recommendation systems (NRS) \cite{yang2024adaptive}, which are critical components of ITS. These systems, like Google Maps or Waze, guide users by providing optimal route recommendations and updating based on real-time traffic conditions. However, they are susceptible to deceptive information-based attacks that compromise both overall system efficiency in terms of travel time and user trust.

The study focuses on misinformed demand attacks, where attackers manipulate the origin and destination (OD) requests submitted to an NRS. By fabricating non-existent demands through techniques like GPS spoofing within the intra-vehicle domain, Sybil attacks, botnets on the inter-vehicle domain, or insider threats compromising the transportation domain, attackers can alter the system's navigational recommendations. This manipulation can misdirect genuine users, causing congestion on targeted roads or benefiting specific entities, such as businesses along those routes.

\begin{figure}
\includegraphics[width=0.7\linewidth]{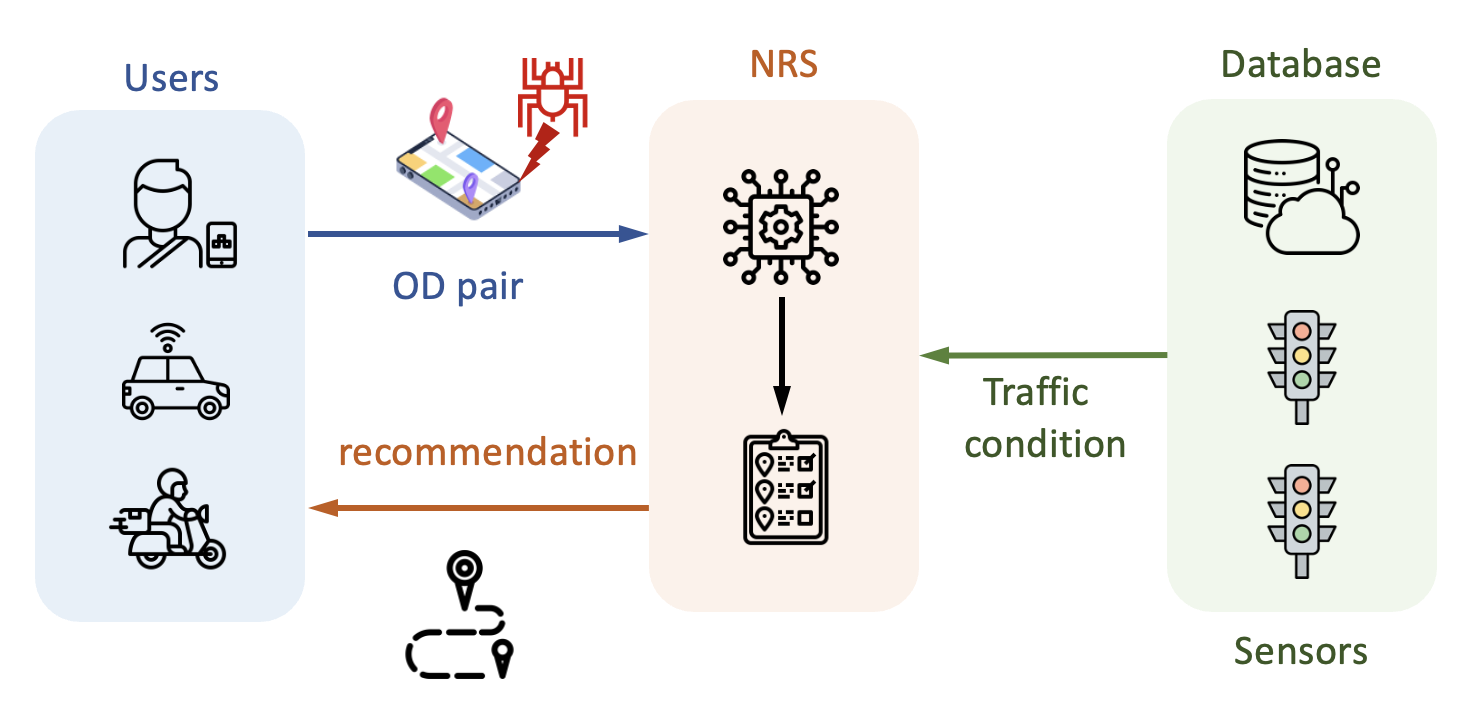}
\centering
\caption{The process for navigational recommendations and where the demand attacks exist.}
\label{fig:NRS}
\end{figure}

For example, attackers might fabricate congestion reports or falsely inflate the demand for certain OD pairs to influence the NRS algorithm, as shown in Fig. \ref{fig:NRS}. This not only misguides users but also destabilizes traffic conditions across the network, leading to inefficiencies and potential safety concerns.

\subsubsection{Game-Theoretic Framework}
PRADA employs a game-theoretic approach to assess and mitigate risks associated with misinformed demand attacks on navigational recommendations and supports the cyber-informed design of resilient systems. It conceptualizes the interaction between attackers and the NRS as a Stackelberg game, a leader-follower model. In this framework, the attacker assumes the role of the leader, strategically fabricating origin-destination (OD) demands to manipulate the system. The NRS acts as the follower, adjusting its route recommendations based on this manipulated data. This interaction forms the foundation for understanding and countering such attacks.

\begin{figure}
\includegraphics[width=0.8\linewidth]{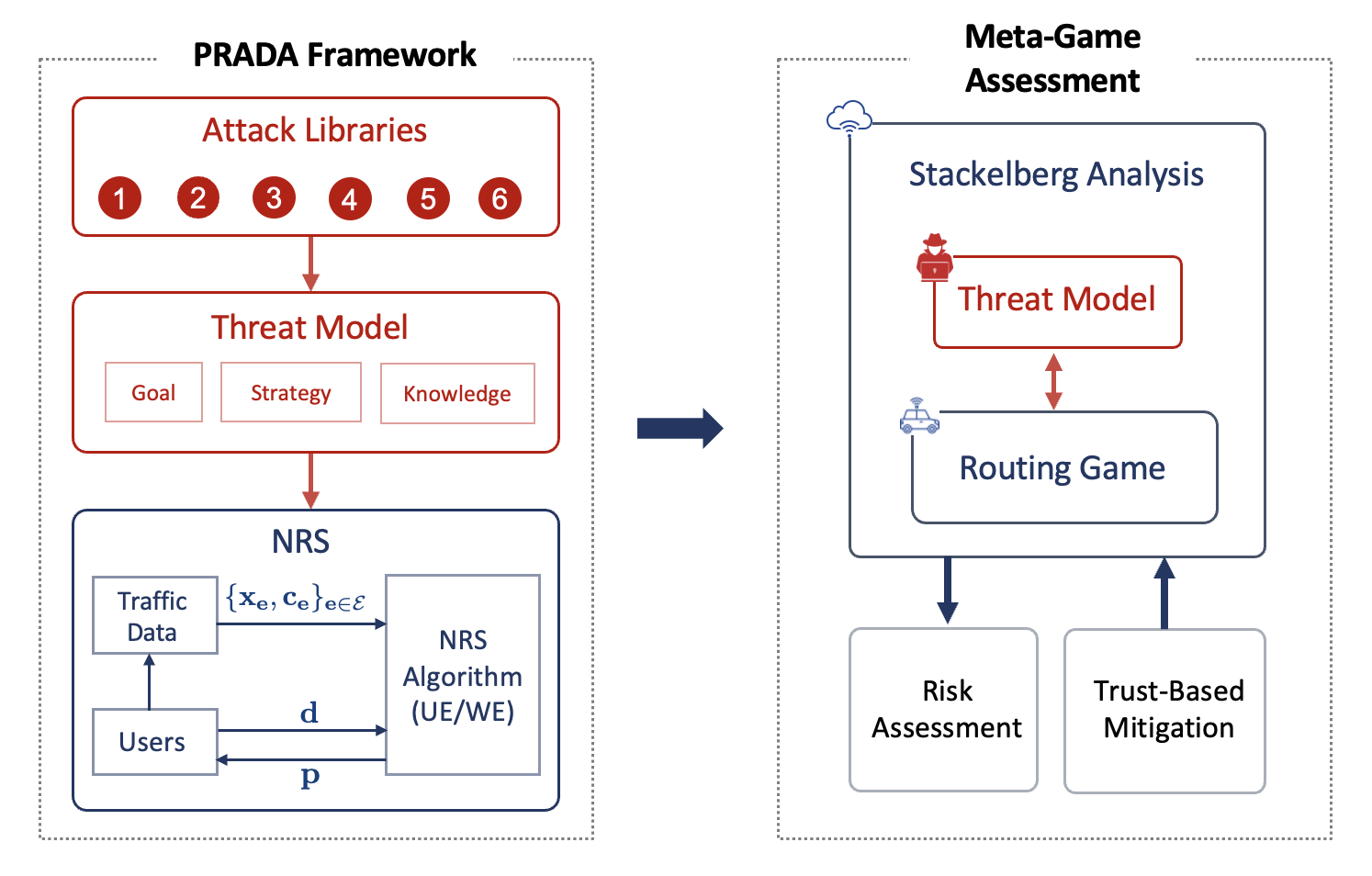}
\centering
\caption{The PRADA framework is analyzed through three layers of games between users, the NRS, the threat model of the attacker, and the PRADA risk evaluator.}
\label{fig:PRADA}
\end{figure}

More specifically, the PRADA framework integrates three analytical layers to proactively and comprehensively measure the risks of possible attacks, as illustrated in Fig. \ref{fig:PRADA}. The first layer is User Equilibrium (UE) Modeling in the routing game, which captures the interaction between users and the NRS. This layer ensures that the system’s recommendations are incentive-compatible, meaning users have no advantage in deviating from the suggested paths. This compatibility fosters user adherence and minimizes unintended behavioral deviations.

The second layer involves Stackelberg Game Analysis, which focuses on modeling the interplay between attackers (threat models) and the NRS. This analysis examines how fabricated demands (or other deceptive informational attacks) influence navigational recommendation decisions, providing insights into the attacker’s strategies and the NRS’s ability to counteract them. By simulating these interactions, the model identifies vulnerabilities and optimal defensive strategies.

The final layer, Meta-Game Assessment, evaluates the broader impacts of various attack scenarios. This layer goes beyond local impacts, considering system-wide effects on traffic flow and overall resilience. By analyzing cascading effects and emergent traffic patterns, it highlights potential risks, mitigation methods, and opportunities for improving system stability.

\paragraph{User Equilibrium (UE) Modeling}
Consider the urban transportation network represented by $\mathcal{G}=\{\mathcal{V}, \mathcal{E}\}$, where the set of nodes $\mathcal{V}$ corresponds to intersections and the set of edges $\mathcal{E}$ indicates the roads. Traveling along a road $e\in \mathcal{E}$ incurs a road-specific cost $c_e: \mathbb{R}_{\ge 0} \mapsto \mathbb{R}_{+}$ associated with the flow $x_e \in \mathbb{R}_{\ge 0}$ on road $e$. One usual choice for the cost $c_e(\cdot)$ is the standard Bureau of Public Roads (BPR) function $$c_e(x_e)=t_e\left(1+\alpha\left(\frac{x_e}{k_e}\right)^\beta\right)$$ for travel time costs. Here, $t_e \in \mathbb{R}_{+}$ represents the free-flow travel time on road $e$, $k_e \in \mathbb{R}_{+}$ signifies the capacity of road $e$, and $\alpha, \beta \in \mathbb{R}_{\ge 0}$ are some parameters.

The user set of NRS is denoted as $\mathcal{U}$. Each user, $u \in \mathcal{U}$, is associated with a specific origin $O_u \in \mathcal{V}$ and destination $D_u \in \mathcal{V}$ pair. We refer to the pair as an OD pair, expressed by $\theta_u = (O_u, D_u)$, and the set of OD pairs for the NRS users is $\Theta_{\mathcal{U}} \subseteq \Theta$, with $\Theta =  |\mathcal{V}| \times |\mathcal{V}|$ representing all the possible OD pairs within the network. Then, user $u$ with OD pair $\theta_u$ has the feasible path choice set $\mathcal{S}_u  = \{s_{u, 1}, \ldots, s_{u, k_u}\}$, which is identical to the feasible path choice set $\mathcal{S}_\theta  = \{s_{\theta, 1}, \ldots, s_{\theta, k_{\theta}}\}$ for OD pair $\theta$, where $\theta=\theta_u$. Each choice $s_{u, i} \in \mathcal{S}_u$ or $s_{\theta, i} \in \mathcal{S}_\theta$ provides the user $u$ a path from the origin to the desired destination. To this end, the elements of the urban transportation network considered by the NRS can be encapsulated using the notation $\mathscr{R}=\left\langle   \mathcal{G}, (c_e(\cdot))_{e \in \mathcal{E}}, \mathcal{U}, (\mathcal{S}_u)_{u \in \mathcal{U}}  \right\rangle$, and we call $\mathscr{R}$ the ``NRS component''.  

Consider the scenario where the NRS recommends a mixed strategy over feasible path choices to the users.
Define $\mathcal{P}_u:=\Delta \mathcal{S}_u$ as the simplex of $\mathcal{S}_u$. A mixed strategy for user $u$ is $\textbf{p}_u \in \mathcal{P}_u$ so that $\textbf{p}_u=\{p_{u, i}\}_{s_{u, i} \in \mathcal{S}_u}$ is a probability distribution over $\mathcal{S}_u$. Each element $p_{u, i}\in [0,1]$ denotes the probability that the NRS recommends path $s_{u, i} \in \mathcal{S}_u$ to user $u$, and needs to satisfy the constraints $\sum_{i=1}^{k_u}p_{u, i}=1 , \ \forall u \in \mathcal{U}$. 
Then, let $\mathcal{P}:=\Pi_{u \in \mathcal{U}}\mathcal{P}_u$, the recommendation suggested by the NRS to all users is $\textbf{p}=\{\textbf{p}_u\}_{u \in \mathcal{U}} \in \mathcal{P}$.

In transportation, from a microscopic perspective, the probability $p_{u, i}$ can be viewed as the expected volume generated by user $u$ along path $s_{u, i}$. This, in turn, contributes to the expected road flow (load) $x_e^r: \mathcal{P} \mapsto \mathbb{R}_{\ge 0}$ on road $e \in \mathcal{E}$ as 
$
    x_e^r(\textbf{p})=\sum_{u \in \mathcal{U}}\sum_{s_{u, i} \in \mathcal{S}_u}p_{u, i}a_{es_{u, i}},
$ where $a_{es_{u, i}}$ is an element of the road-path incidence matrix $A_{|\mathcal{E}|\times |\Pi_{u \in \mathcal{U}} \mathcal{S}_u|}=[a_{es_{u, i}}]$, and is defined as follows.
$$ a_{es_{u, i}}=
\begin{cases}
    1 \qquad \text{if} \ e \in s_{u, i},\\
    0 \qquad \text{otherwise}.
\end{cases}
$$ Hence, a generalized travel cost $C_{u, i}: \mathcal{P} \mapsto \mathbb{R}_{+}$ for path $s_{u, i}$ can be formulated by summing the costs of all the roads along the path:
$
    C_{u, i}(\textbf{p})=\sum_{e \in s_{u, i}}c_e(x_e^r(\textbf{p})).
$ In this context, the expected cost evaluated by user $u$ is $F_u^r: \mathcal{P} \mapsto \mathbb{R}_{\ge 0}$, where
\begin{equation}
    F_u^r(\textbf{p}_{u},\textbf{p}_{-u}) = \sum^{k_u}_{i=1}p_{u,i}C_{u,i}(\textbf{p}_{u},\textbf{p}_{-u}).
\label{eq:F_u}
\end{equation}
Note that a recommendation $\textbf{p} \in \mathcal{P}$ can be interpreted as the strategy profile in a routing game between NRS users. Hence, the routing game addressed by the NRS can be defined as $\Gamma^r=\langle \mathscr{R}, \mathscr{F}^r \rangle$, where $\mathscr{F}^r=(F_u^r)_{u \in \mathcal{U}}$ represents the costs evaluated by users.
However, human users may choose not to follow the NRS recommendation if they find a better alternative. Therefore, 
the NRS must ensure that given the recommendations \(\textbf{p}_{-u}\) to users other than \(u\), user \(u\) has no incentive to unilaterally deviate from the recommended \(\textbf{p}_u\). This coincides with the concept of user equilibrium (UE), which is defined as follows:
\begin{definition}[User Equilibrium Recommendation]
    Considering a routing game addressed by the NRS defined as $\Gamma^r=\langle\mathscr{R}, \mathscr{F}^r\rangle$, a mixed strategy profile $\textbf{p} \in \mathcal{P}$ for all the users is called a user equilibrium recommendation if it satisfies:
    \begin{align}
    F_u^r(\textbf{p}_{u},\textbf{p}_{-u})-F_u^r(\textbf{p}^{\prime}_{u},\textbf{p}_{-u}) \leq 0 , \forall \ \textbf{p}^{\prime}_{u} \in \mathcal{P}_u, \forall u \in \mathcal{U}.
    \label{prob:RS}
    \end{align}
\label{def:NRS}
\end{definition}
UE recommendation can be found by gradient descent-based method \cite{yang2024adaptive}. Let $\text{proj}_{\mathcal{P}_u}$ represent the projection onto simplex $\mathcal{P}_u$ and $\rho \in \mathbb{R}$ denote the step size, problem \eqref{prob:RS} can be solved by finding a fixed point to:
\begin{equation}
    \textbf{p}_u^{*} = \text{proj}_{\mathcal{P}_u} \left[\textbf{p}_u^{*} - \rho \nabla_u F_u^r(\textbf{p}_u^{*}, \textbf{p}_{-u}^{*})\right], \ \forall u \in \mathcal{U}.
\label{eq:PGD}
\end{equation} 

\paragraph{Stackelberg Game Analysis}
Recall that $\mathcal{U}$ represents the user set of the NRS, with their associated set of OD pairs, denoted as $\Theta_{\mathcal{U}}$. For each OD pair $\theta \in \Theta_{\mathcal{U}}$, the demand flow aggregated from users that must be routed from the corresponding origin to the desired destination is $d_\theta = \sum_{u \in \mathcal{U}} \mathbf{1}_{\{\theta_u = \theta\}}$. For OD pair $\theta \in \Theta \setminus \Theta_{\mathcal{U}}, d_\theta = 0$. As we consider the case of deceptive misinformed demand attacks, let vector $\textbf{d}:=\{d_\theta\}_{\theta \in \Theta}$ and denote the set of solutions to NRS's problem defined in Definition \ref{def:NRS} associated with demand $\textbf{d}$ as $U(\textbf{d})$.

Consider an example of strategic attackers with local-targeted attack objectives. Specifically, the attacker intends to have a desired level of expected flow load caused by genuine NRS users on the target road $e^{\prime} \in \mathcal{E}$. That is, the attacker aims to make $x_{e^\prime}^r(\textbf{p}) = \sum_{u \in \mathcal{U}}\sum_{s_{u, i} \in \mathcal{S}_u}p_{u, i}a_{e^\prime s_{u, i}}$ in UE recommendation achieve a desired level $\gamma \in \mathbb{R}_{\ge 0}$.

In the case where a Sybil-based attacker generates non-existent demands $\textbf{d}^a \in \mathcal{D}$, where $\mathcal{D}=\mathbb{Z}_{\ge 0}^{|\Theta|}$ using Sybil (fake) users so that the user set considered by the NRS becomes $\mathcal{U}^\prime$. Then, the NRS will need to consider an aggregated demand of $\textbf{d}^{\prime}=\textbf{d} + \textbf{d}^a$ when generating the recommendation. Note that for each OD pair $\theta$, the demand $d_\theta^{\prime}$ under attack consists of $d_\theta + d_\theta^a$. Without loss of generality, we can assume that only a proportion of $\frac{d_\theta}{d_\theta + d_\theta^a}$ of the expected path flow (for OD pair $\theta$) $\sum_{u \in \mathcal{U}^\prime}p^{\prime*}_{u, i}\mathbf{1}_{\{\theta_u = \theta\}}$ with respect to solution $\textbf{p}^{\prime *} \in U(\textbf{d}^{\prime})$ is caused by authentic users. In this case, the attacker aims to make the flow load from genuine users, denoted as $x_{e^\prime}^u(\textbf{p}^{\prime *}) = \sum_{\theta \in \Theta}\sum_{u \in \mathcal{U}^\prime}\sum_{s_{u, i} \in \mathcal{S}_u}\frac{d_\theta}{d_\theta + d_\theta^a}\mathbf{1}_{\{\theta_u = \theta\}}p^{\prime*}_{u, i}a^\prime_{e^\prime s_{u, i}}$,  on the targeted road $e^\prime$ reach the desired level $\gamma$. To this end, the Stackelberg game between the attacker (AT) and the NRS can be defined as $\Gamma^s=\langle \text{AT}, \mathcal{D}, U_{AT}, (e^\prime, \gamma), \Gamma^r \rangle$, where $U_{AT}: \mathcal{D} \mapsto \mathbb{R}_{\ge 0}$ is the attacker's cost in terms of the resources spent in fabricating fake demands. The leader-follower problem is formulated as follows and can be solved by gradient descent-based algorithms \cite{boyd2004convex}.
\begin{subequations}
    \begin{align}
        \min_{\textbf{d}^a} \ & U_{AT}(\textbf{d}^a)=\sum_{\theta \in\Theta} d_\theta^a\\
        \text{s.t.} \ &x_{e^\prime}^u(\textbf{p}^{\prime *}) \geq \gamma, \\
        &\textbf{p}^{\prime *} \in U(\textbf{d} + \textbf{d}^a), \\
        & d_\theta^a \geq 0, \forall \theta \in \Theta.
    \end{align}
\label{prob:RS_attack}
\end{subequations}

\paragraph{Meta-Game Assessment} To assess the risks, we introduce two metrics as the outcomes of our PRADA framework. 
Let $\textbf{p}$ be the recommendation to all the users without attack and $\textbf{p}^\prime$ is the one under attack.

The \textit{targeted impact metric (TI)} is defined as the difference in traffic flow on each road with and without the demand attack, divided by the flow without the attack. Specifically, for each road $e \in \mathcal{E}$, the measure $TI_{e}$ is given by:
\begin{equation}
TI_e = \frac{|x_e^r(\textbf{p}^\prime) - x_e^r(\textbf{p})|}{x_e^r(\textbf{p})},  
    \label{eq:ti}
\end{equation} which can assist in measuring the percentage change in traffic flow on the specific or targeted road affected by the demand attack. A larger value of $TI_e$ indicates the road $e$ is influenced more, often implying higher risk under the attack.

Given the metrics $TI_{e}, \forall e \in \mathcal{E}$, the \textit{network impact metric (NI)} is then defined as the mean of $TI_{e}$ across all the roads/edges within the network. The measure $NI$ is as follows:
\begin{equation}
NI = \frac{1}{|\mathcal{E}|}\sum_{e \in \mathcal{E}} TI_{e}=\frac{1}{|\mathcal{E}|}\sum_{e \in \mathcal{E}} \frac{|x_e^r(\textbf{p}^\prime) - x_e^r(\textbf{p})|}{x_e^r(\textbf{p})}. 
    \label{eq:ni}
\end{equation} The metric $NI$ allows us to evaluate the percentage change in traffic flow across the entire network. It is important to note that if the demand attack primarily affects traffic flow on roads within a small area, as indicated by the $TI_{e}$ values for those roads, this localized impact will be averaged out when considering the network-wide impact.

Then, the PRADA risk evaluator can measure the risk for different threat models from different types of attacks in the attack libraries, compare the targeted and network impacts, and design methods for defense or mitigation purposes. For example, if the risk evaluator aims to assess the risk considering the scenario where the attacker with local-targeted objectives performs misinformed demand attacks, the risk evaluator can adjust parameters of $\Gamma^s$ in the Stackelberg Game Analysis layer and get insights from the framework outputs, TI and NI.

\subsubsection{Key Insights from the Case Study}
\begin{figure}
\includegraphics[width=0.95\linewidth]{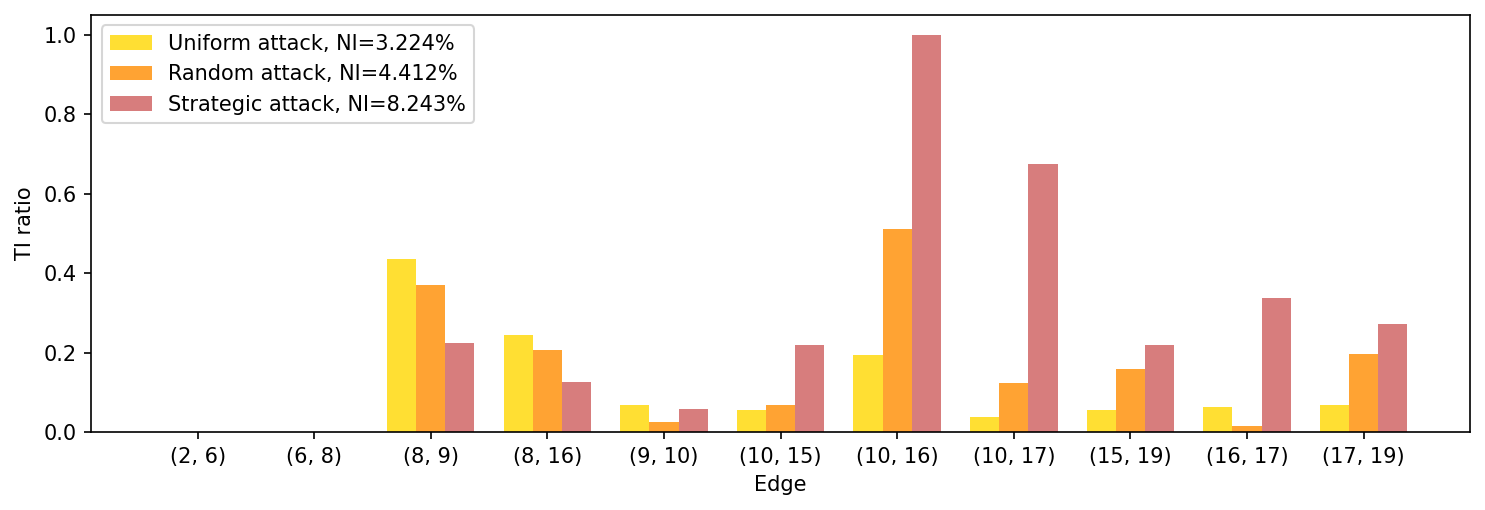}
\centering
\caption{Risk report for misinformed demand attacks targeted edge $(10, 17)$ in terms of TI (local-targeted impact on roads along users' feasible paths) and NI (network-wide impact) when encountering non-strategic (random, uniform) and strategic attackers. Here, strategic attackers refer to the ones who solve Problem \ref{prob:RS_attack} while non-strategic ones manipulate demands by uniformly or randomly increasing or decreasing the number of requests associated with some OD pairs.}
\label{fig:exp1_}
\end{figure}

Simulation experiments conducted under the PRADA framework reveal several critical insights. One notable finding is the impact of local-targeted attacks, where attackers redirect user flow to specific roads to benefit certain entities, such as businesses or political interests. For example, attackers might manipulate OD requests to artificially increase traffic through a specific area, creating commercial advantages or political leverage. These localized manipulations demonstrate how small-scale attacks can disproportionately affect user behavior and traffic patterns. Fig. \ref{fig:exp1_} shows the risk report in terms of network-wide impact (NI) and local-targeted impact (TI) on each road/edge when the attacker targets edge $(10, 17)$, and we can observe both strategic and non-strategic indeed affect genuine user's behavior and traffic patterns.

Another key insight is the system-wide risks posed by such attacks. Even localized disruptions can have cascading effects, destabilizing broader traffic networks. For instance, in Fig. \ref{fig:exp1_}, a locally targeted attack for edge $(10, 17)$ eventually leads to network-wide impact indicated by NI in the figure, and the risks posed by strategic attackers are higher compared to non-strategic ones. 

\begin{figure}
\includegraphics[width=0.75\linewidth]{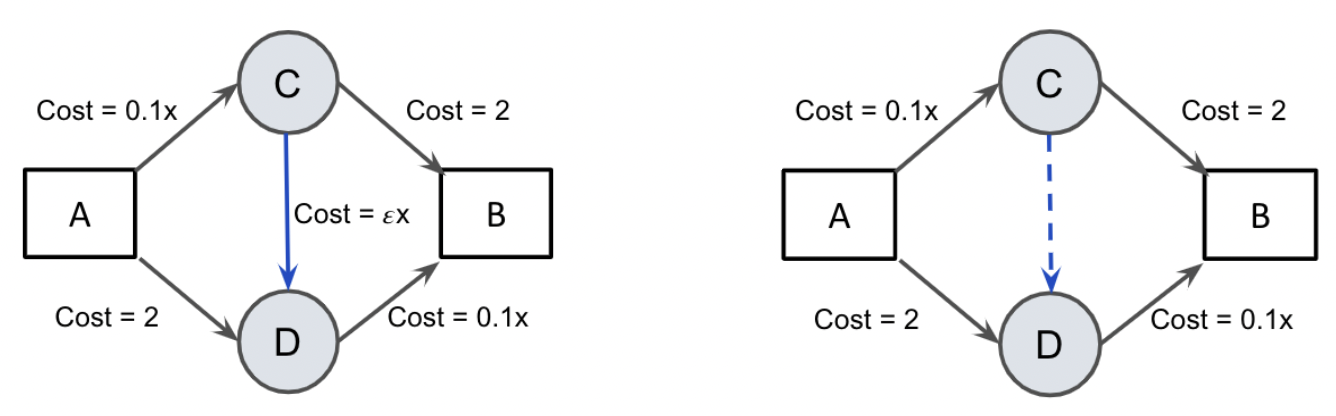}
\centering
\caption{A carefully crafted example for the discussion on the ``Fragile Paradox''. Left: without attack. Right: with an attack.}
\label{fig:paradox}
\end{figure}

Interestingly, the study also uncovers a ``Fragile Paradox'', where some attacks inadvertently improve traffic conditions by redistributing flow across underutilized routes. This paradox highlights the complexity of ITS dynamics and underscores the importance of proactive risk assessment to anticipate both intended and unintended consequences. We use the transportation network shown in Fig. \ref{fig:paradox} as an illustrative example. There are $30$ users aiming to go from node A to node B, and the $\epsilon$ is small enough so that the cost on C-D is close to $0$ even though all $30$ users are passing through. Before the attack, the NRS will recommend a mixed strategy $(1/3, 1/3, 1/3)$ on paths A-C-B, A-C-D-B, and A-D-B, respectively. The overall costs on these three paths are all $4$, which leads to a total travel time cost of $120$ for users. Suppose the attacker wants more `` genuine users'' to pass D-B by fabricating a large demand on C-D to make C-D seem congested to the NRS. The NRS will recommend a strategy $(0.5, 0, 0.5)$ on paths A-C-B, A-C-D-B, and A-D-B, respectively. The overall costs on A-C-B and A-D-B are both $3.5$, which leads to the total travel time cost for users becoming $105$. Therefore, we can conclude that the cost under attack is ``better'' than the performance without attack in this carefully crafted example.

\subsubsection{Mitigation Strategies}
To address the vulnerabilities exposed by these insights, PRADA introduces a trust mechanism to enhance the resilience of NRS against demand attacks. Central to this mechanism are trust scores, \( T_u \in \mathbb{R}_{\ge 0} \) for all \( u \in \mathcal{U} \), which quantify user confidence in the system’s recommendations. Based on the trust scores, PRADA emphasizes trusted recommendations, which integrate trust constraints defined below directly into the NRS’s recommendation algorithms.

\begin{definition}[Trust Constraint (TC)]
    Consider an NRS component, denoted as $\mathscr{R}$. A recommended mixed strategy $\textbf{p}_u \in \mathcal{P}_u$ for a user $u$ is said to satisfy the trust constraint if the distance (in terms of Kullback–Leibler (KL) divergence) between the currently and previous recommended strategy $\textbf{p}_u^o \in \mathcal{P}_u$ is less than the trust score $T_u \in \mathbb{R}_{\ge 0}:$
    \begin{equation} D(\textbf{p}_u||\textbf{p}_u^o)=\sum^{k_u}_{i=1} p_{u,i} \log \left(\frac{p_{u,i}}{p_{u,i}^o}\right) \leq T_u.
    \label{eq:TC}
    \end{equation}
\label{def:TC}
\end{definition}

The trust constraints act as a regulatory mechanism, limiting how much the NRS recommendation can deviate from historical traffic patterns. Higher trust scores enable more adaptive and flexible recommendations, while lower scores prioritize consistency, reducing the system's susceptibility to manipulation. In this context, the trusted recommendation ensures that current recommendations align closely with user expectations and historical traffic conditions, thereby enhancing user adherence and the overall reliability of the system. By balancing flexibility and consistency, these mechanisms help mitigate the impact of demand attacks while maintaining the integrity of traffic management.

\section{Conclusions and Discussions}

This chapter introduces the concept of deceptive information attacks within Intelligent Transportation Systems (ITS), focusing on their distinctive characteristics and critical implications. These attacks represent an emerging class of threats that exploit the interconnectedness and interdependencies of ITS, with impacts that span multiple domains. Unlike traditional attacks, such as malware or ransomware, which often target individual devices or systems locally, deceptive information attacks propagate through the network, leveraging the flow of information to achieve widespread disruption.

The deceptive nature of these attacks allows adversaries to manipulate not only the systems but also human operators and users, tricking them into taking actions that serve the attacker’s goals. For example, attackers may use falsified traffic updates to misdirect vehicles or induce panic among users, leading to systemic inefficiencies or safety hazards. These attacks exploit vulnerabilities across different domains, such as intra-vehicle, inter-vehicle, transportation infrastructure, and human interactions, creating cascading effects that amplify their impact.

Furthermore, many of these attacks emerge from the composition of multiple attack vectors that cross domain boundaries \cite{ge2024mega,chen2019games,chen2019control,zhu2015game}. For instance, a phishing attack targeting an operator (human domain) can introduce malware into the transportation infrastructure (system domain), which then propagates to vehicles (intra-vehicle domain), causing widespread disruption. Mapping out these attack vectors and understanding their cross-domain interdependencies is essential for managing their risks. This requires a holistic, cross-domain approach to identify, mitigate, and respond to such threats effectively.

One example of this complexity is misinformed demand attacks on navigational systems, where attackers fabricate false traffic data to misdirect users and disrupt traffic flow. These attacks demonstrate how deceptive information can exploit system interconnectivity, manipulating user behavior and creating cascading failures across the ITS ecosystem. Addressing these challenges demands integrated strategies, combining domain-specific expertise with cross-domain resilience mechanisms, to protect ITS from the evolving landscape of cyber threats.

Dealing with such attacks is inherently challenging due to their complexity and cross-domain nature. There is a pressing need for a comprehensive framework that provides a clear view of the possible attacks within each domain while also enabling an understanding of how these attacks interact to create system-wide issues. This dual perspective—domain-specific insights and cross-domain integration—is crucial for identifying vulnerabilities and crafting effective mitigation strategies.
A holistic understanding of the threats is essential for developing holistic solutions. Such solutions must be designed to operate effectively at a local level within each domain while simultaneously contributing to the overall resilience of the ITS. By aligning local defense mechanisms with system-wide objectives, ITS can ensure coordinated and adaptive responses to threats, minimizing their impact and preventing cascading failures.

Game theory provides a powerful framework for addressing the challenges of securing ITS. Its inherent ability to model strategic interactions among multiple agents across domains makes it particularly well-suited for assessing risks holistically. By piecing together the behaviors and objectives of agents—such as users, system designers, and attackers—game-theoretic models enable a comprehensive evaluation of risks that accounts for both localized impacts and system-wide interdependencies.
The game-theoretic models developed within this framework allow for not only risk assessment but also the design of strategies to mitigate those risks. These strategies can be implemented in a decentralized manner, leveraging the multi-agent structure of the models. Such decentralization is critical for scalable and adaptive solutions, particularly in the highly interconnected and dynamic context of ITS.

Dynamic games, when integrated with control and learning theories, add another dimension of flexibility and robustness. They enable adaptive and non-myopic solutions that plan over long time horizons, incorporating predictive elements to anticipate and counteract evolving threats. This combination ensures that the solutions remain effective even in the face of unforeseen challenges or changes in the threat landscape.

PRADA case study exemplifies this approach. PRADA begins by modeling the interactions among key agents: the users, who navigate the transportation system; the system designer, who aims to optimize system performance and safety; and the attackers, who seek to disrupt or exploit the system. These agents are treated as players in a Stackelberg game, where the equilibrium solution allows the prediction of potential risks arising from deceptive information attacks.

The framework incorporates a trust component to influence risk levels dynamically. By assigning and updating trust scores based on user behavior and system reliability, PRADA can adjust recommendations and mitigate the impacts of attacks. Furthermore, a dynamic adaptive learning process enhances the system's resilience by enabling it to learn from past incidents and adapt to new threats in real-time.
This research methodology is highly adaptable and can be extended to address a wide range of attacks and design solutions within ITS. For instance, it can be applied to investigate spoofing attacks in ITS \cite{zhang2017strategic,chen2019control}. In such attacks, adversaries manipulate communication protocols or physical signals to mislead system components, such as sending falsified GPS signals to vehicles or tampering with V2V  and V2I communication. By modeling these interactions using game-theoretic frameworks, we can identify the strategies attackers use to maximize disruption and design countermeasures that anticipate and mitigate their impacts. These countermeasures could include adaptive filtering, trust-based communication protocols, and real-time anomaly detection systems.

Similarly, the methodology can be used to study APTs  in ITS. APTs are sophisticated, long-term attacks often orchestrated by well-resourced adversaries, such as nation-states or organized crime groups \cite{pawlick2017strategic,huang2020dynamic,zhu2018multi,huang2019dynamic,huang2019adaptive}. These attackers exploit system vulnerabilities to gain unauthorized access, persist within the network undetected, and execute targeted actions, such as disrupting traffic control centers or compromising critical infrastructure. Game-theoretic models can help map the interactions between defenders and APT actors, capturing the stealthy and strategic nature of these attacks. By integrating dynamic game approaches with learning algorithms, the system can adapt its defenses, such as deploying decoys, tightening access controls, and reinforcing endpoint security, to proactively counteract APTs.

Beyond spoofing and APTs, this methodology can be extended to man-in-the-middle attacks, data poisoning, DoS attacks, and ransomware, among others. Each attack type can be modeled with specific adversarial strategies and system vulnerabilities, enabling tailored defense mechanisms. For example, for DoS attacks, dynamic resource allocation and load-balancing strategies could be optimized through game-theoretic analysis, ensuring that critical services remain operational even under attack. Similarly, for ransomware, predictive models can be developed to identify early indicators of an attack and isolate affected systems. 
In all these cases, the combination of game theory, control mechanisms, and adaptive learning provides a robust framework for analyzing and addressing the challenges posed by cyber threats in ITS. This systematic approach ensures that solutions are not only effective against current threats but are also resilient and adaptive to future challenges, making it a versatile tool for enhancing cybersecurity across the ITS ecosystem.

\bibliographystyle{splncs04}
\bibliography{mybibliography}
\end{document}